\documentstyle[aps]{revtex}

\begin{document}

% \draft command makes pacs numbers print
\draft
\preprint{UAM-GA-29}
\quad
\newcommand{\noi}{\noindent}
\newcommand{\be}{\begin{equation}}
\newcommand{\ee}{\end{equation}}
\newcommand{\bea}{\begin{eqnarray}}
\newcommand{\eea}{\end{eqnarray}}
\newcommand{\ba}{\begin{array}}
\newcommand{\ea}{\end{array}}

%%%% SYMBOLS %%%%%%

\newcommand{\ie}{{\it i.e.}}
\newcommand{\etal}{{\it et.al.}}
\newcommand{\eg}{{\it e.g.}}

% repeat the \author\address pair as needed
\title{ELECTROMAGNETIC FIELD IN SOME ANISOTROPIC STIFF FLUID UNIVERSES }
\author{Luis O. Pimentel }
\address{ Departamento de F\'{\i}sica,\\
 Universidad Aut\'onoma Metropolitana-Iztapalapa,
Apartado Postal 55-534, M\'exico D. F., MEXICO.}
\date{\today}
\maketitle
\begin{abstract}
% insert abstract here
The electromagnetic field is studied in a family of exact solutions of
the Einstein equations whose material content is a perfect fluid with
stiff equation of state (p = $\epsilon $ ). The field equations
are solved exactly for several members of the family.

\end{abstract}
% insert suggested PACS numbers in braces on next line
to appear in Modern Physics Letters A

% body of paper here

% now the references. delete or change fake bibitem. delete next three
%   lines and directly read in your .bbl file if you use bibtex.
%\begin{references}
%\bibitem{tag} Fake bibitem.
%\end{references}

% figures follow here
%
% Here is an example of the general form of a figure:
% Fill in the caption in the braces of the \caption{} command. Put the label
% that you will use with \ref{} command in the braces of the \label{} command.
%
% \begin{figure}
% \caption{}
% \label{}
% \end{figure}

% tables follow here
%
% Here is an example of the general form of a table:
% Fill in the caption in the braces of the \caption{} command. Put the label
% that you will use with \ref{} command in the braces of the \label{} command.
% Insert the column specifiers (l, r, c, d, etc.) in the empty braces of the
% \begin{tabular}{} command.
%
% \begin{table}
% \caption{}
% \label{}
% \begin{tabular}{}
% \end{tabular}
% \end{table}

%\end{document}
%
% ****** End of file template.aps ******

%\font \gross=ambx10 scaled\magstep2
%	\def\sqr#1#2{{\vcenter{\vbox{\hrule height.#2pt
%	\hbox{\vrule width.#2pt height#1pt \kern#1pt
%	   \vrule width.#2pt}
%	\hrule height.#2pt}}}}

\def\square{\mathchoice\sqr54\sqr54\sqr{6.1}3\sqr{1.5}6\,\,}
\def\qa {^{\prime}}
\def\qq {^{\prime\prime}}
\def\qb {^{\prime 2}}
\def\noi {\noindent}
%\baselineskip=20pt
%\magnification=1200 \vglue0pt plus1fil

\section { Introduction}

Recently Lotze ${}^1$, using a method developed by Sagnotti and Zwiebach
${}^2$,
 has considered the Maxwell equations in anisotropic universes with a
 diagonal Bianchi type I metric and found exact solutions for two particular
 expansions laws in the axisymmetric case. Electromagnetic fields
in space-times with local rotational symmetry, using the Debye-
formalism were also considered by Dhurandhar et al. ${}^3$; they
obtained exact solutions for Kantowski-Sachs universes, Taub space
and for Bianchi type I with dust as material content. Here I consider the
same problem
in a family of Bianchi type I models that are solutions of
Einstein equation with either a free massless scalar field or with a
perfect fluid with stiff equation of state. The metric was obtained by Jacobs
${}^4$
and is also given by Vajk and Eltgroth
 ${}^5$ and it is  a particular case of the metrics studied by Thorne ${}^6$,
more
recently it was rediscovered by Iyer and Vishveshwara ${}^7$ while looking for
exact solutions in which Dirac equation separates. Recently, the production of
scalar
particles in these model was considered ${}^9$. We write the metric in the
following form,
\be
ds^2 =-dt^2+ t^{2q} (dx^2 +dy^2 )+t^{2(1-2q)} dz^2,
\ee
This metric is a one parameter family of  solutions to Einstein equations
 with a perfect stiff fluid. The parameter q is related to the
Lagrangian of the scalar field or to the energy density of the
perfect fluid by
the relation
\be
L=(1/2)\phi^{;a}\phi_{;a}={q(2-3q)\over t^2}.
\ee
or

\be
\epsilon =p ={q(2-3q)\over t^2}.
\ee

This metric is also the solution to Einstein equations with a massless
minimally coupled scalar field.
 The qualitative features of the expansion depend on q in the following way:
for $1/2 < q $, the universe expands from a "cigar" singularity; for
$q=1/2$, the universe expands purely transversely from an initial "barrel"
singularity; for $0<q<1/2 $, the initial singularity is "point"-like; if
$q\leq
0$ we have a "pancake" singularity. The case
$q=1/3$ is the isotropic universe with a stiff fluid; the case p=q is the
Minkowski spacetime. This family of metrics is "Kasner-like" in the sense that
the sum of the exponents is equal to one but the sum of the squares is not
equal to one except in the two cases when $q=0$ and $q=2/3$ when we have
vacuum. The symmetries of these spacetimes can be described by four spacelike
Killing vector fields,
\be
\xi _1 =x {\partial \over \partial y} -y{\partial \over \partial x},\quad
\xi _2
= {\partial \over \partial x},\quad  \xi _3 = {\partial \over \partial y},
\quad
  \xi _4 = {\partial \over \partial z},
\ee

The first vector corresponds to the rotational simmetry in the plane xy and
the
other three to the translational simmetry along the x, y and z axis. The
non-vanishing commutators are
\be
[\xi _1 , \xi _2 ] = \xi _3,[\xi _3 , \xi _1 ] =- \xi _2 .
\ee

\noi In the next section we review the formalism used by Sagnotti and
Zwiebach ${}^2$, that write the metric in the following form,

\be
ds^2 =-C^2 d\tau ^2+ C_1^{2 } (dx^2 +dy^2 )+C_3^{2 } dz^2,
\ee

\noi comparing with Eq.(1) we see that in our particular case we
have

\be
\tau = t;C=1,\quad C_1 =t^{q},\quad C_3=C^{1-2q} .
\ee

{\sl Notations and conventions}. $c=1$, metric signature (-+++); greek
indices run from 0 to 3 , latin indices from 1 to 3. The derivative with
respect to the time $\tau$ (or t) is denoted by an overdot.

\section {FIELD EQUATIONS}

 The Maxwell equations will be written in the following the method developed by
Sagnotti and Zwiebach ${}^2$ . The field strength tensor $F^{\mu \nu} $ is
written
as

\be
F^{\mu \nu} \propto \int d^3 {bf k}f^{\mu \nu} ({\bf k};\tau )\exp (i{\bf kx} )
{}.
\ee
\noi The Maxwell equation are written in terms of the quantities

\be
F^{m(\sigma)} =\sqrt{-g} (f^{0m} +i \sigma \;\,^* f^{m0}),\quad\sigma =\pm ,
\ee

\noi using spherical coordinates in ${\bf k}$ space, only the  $\theta$ and
$\phi$ and componets of $F^{m(\sigma)}$ are nonvanishing and will be denoted
by $F^{(\sigma)}$ and $G^{(\sigma)}$, respectively, satisfying the equations

\be
{\dot F}^{(\sigma)} =-\sigma k [\alpha { F}^{(\sigma)} +\beta {
G}^{(\sigma)}]
\ee
\noi and

\be
{\dot G}^{(\sigma)} = =\sigma k [\gamma { F}^{(\sigma)} +\alpha  {
G}^{(\sigma)}]
\ee

\noi with

\be
\alpha = {C^2 \over\sqrt{ -g}}{k_1k_2k_3\over k {k_{\perp}}
^2}(C_2^2-C_1^2),
\ee

\be
\beta = {C^2 \over\sqrt{ -g}}{1\over  {k_{\perp}}
^2}(C_1^2 k_2^2+C_2^2 k_1^2)
\ee

\noi The parameter $\gamma$ is given by $\beta \gamma -\alpha ^2 =C^2
\Omega ^2
/k^2$, where

\be
\Omega ^2 = \sum (k_i /C_i)^2 .
\ee

\noi After the elimination of ${G}^{(\sigma)})$ the relevant field equation is

\be
{\ddot F}^{(\sigma)} - {{\dot \beta}\over \beta}
{\dot F}^{(\sigma)}+[C^2 \Omega ^2 + \sigma k \beta
{({\alpha \over \beta })^{\; .} \;  }\; ]
{\dot F}^{(\sigma )} =0.
\ee

\noi In the present case we have $\alpha =0$ and

\be
\beta = (b\tau )^{3q-1},\qquad \Omega ^2 = k_\perp ^2 (b\tau )^{-3q} +
k_3 ^2
(b\tau )^{6q-3}
\ee

\be
 {\ddot F}^{(\sigma)} +{(1-3q)  \over \tau}
{\dot F}^{(\sigma)}+[ k_\perp ^2 (b\tau )^{1-3q} +k_3 ^2
(b\tau )^{6q-2}]F^{(\sigma )}=0.
\ee

Because we are considering axisymmetric case the solutions to the field
equation
become independent of the polarisation $\sigma $. There are several cases
where
the above equation can be solved exactly and are considered in the following
section.

\section {EXACT SOLUTIONS}

In this section we consider those values of q for which it is possible to
solve
equation (10) for arbitrary values of $k_3$ and $k_\perp$.

\subsection {q=0.}

In this case the field equation is

\be
 {\ddot F}^{(\sigma)} +
{{\dot F}^{(\sigma)}\over \tau }+
[ k_\perp ^2 (b\tau ) +
{k_3 ^2 \over (b \tau)^2} ]    F^{(\sigma )}=0,
\ee

\noi with the solution

\be
F^{(\sigma )} =c_1 H_\nu ^{(1)}(|k_\perp | (b\tau )^{3/2})+
	      c_2 H_\nu ^{(2)}(|k_\perp | (b\tau )^{3/2}),
\ee

\noi where $\nu  = i|k_3 |/2$ and $H_\nu ^{(i)} $ is a Hankel function of
order
$\nu$ and $c_i$ are integration constants.

\subsection {q=1/5.}

For this value of q the field equation is

\be
 {\ddot F}^{(\sigma)} +{2  \over 5}
{{\dot F}^{(\sigma)}\over \tau }+[ k_\perp ^2 (b\tau )^{2/ 5} +k_3 ^2
(b\tau )^{-{4/ 5}}]F^{(\sigma )}=0
\ee

\noi and the solution is given by

\be
F^{(\sigma )}=c_1 D_{a} (\eta ) +
c_2 D_{-(a+1)} (i\eta ),
\ee

\noi here $D_a$ is the parabolic function of order a with

\be
\eta =\pm (1+i) \sqrt{5k_\perp /2} (b \tau )^{3/5} ,
\ee

\noi and

\be
 a = -{1\over 2}-{5k_3^2\over 4k_\perp }.
\ee

\subsection {q=1/4.}

Now the field equation is

\be
 {\ddot F}^{(\sigma)} +{1\over 4 \tau}
{\dot F}^{(\sigma)}+[ k_\perp ^2 (b\tau )^{1\over 4} +k_3 ^2
(b\tau )^{-1\over  2}]F^{(\sigma )}=0,
\ee

\noi and the solution is as follows

\be
F^{(\sigma )}=\sqrt{ \eta }[c_1 H_{1/3} ^{(1)} ({2\over 3} {\eta}^{3/2} ) +
c_2 H_{1/3} ^{(2)} ({2\over 3} {\eta}^{3/2} )],
\ee

\noi here $H_\nu ^{(i)} $ is a Hankel function of order
$\nu$ with

\be
\eta = {\kappa (b\tau )^{3/4} + \lambda \over {\kappa}^{2/3}} ,
\ee

\noi and

\be
\kappa = 4k_\perp ^2, \qquad \lambda = 4 k_3^2.
\ee

\subsection{  q=1/3.}

This case is the isotropic Robertson-Walker with a stiff fluid and a $t^{1/3}$
expansion law. The field
equation is

\be
\ddot F^{(\sigma )} +k^2 F^{(\sigma )} =0,
\ee

\noi with

\be
    k^2 =    k_\perp ^2 + k_3 ^2,
\ee

\noi and the solutions is,

\be
F^{(\sigma )}=
	    c_1 \exp(ik\tau )+ c_2 \exp(-ik\tau ).
\ee

\subsection{ q=1/2.}

\be
 {\ddot F}^{(\sigma)} -
{{\dot F}^{(\sigma)}\over2 \tau }+
[ k_3 ^2 (b\tau ) +
{k_\perp ^2 \over \sqrt{(b \tau)}} ]  F^{(\sigma )}=0,
\ee

\noi with the solution

\be
F^{(\sigma )}= c_1 F_0 (\eta ,\rho )+ c_2 G_0 (\eta ,\rho ),
\ee

\noi where $ F_0 (\eta ,\rho )$and $ G_0 (\eta ,\rho )$ are the regular and
the
irregular (logaritmic) Coulomb wave functions ${}^7$ with null angular momentum
and

\be
\eta  =- ({k_\perp ^2 \over 2 k_3 ^2})\qquad  {\rm and} \qquad \rho = (b
\tau
^{2/3} ).
\ee

\subsection {q=1.}

The equation (10) is in this case

\be
{\ddot F}^{(\sigma)} -  2
{{\dot F}^{(\sigma)}\over \tau }+
[ k_\perp ^2 (b\tau )^{-2} +
{k_3 ^2 {(b \tau)}^4} ]    F^{(\sigma )}=0,
\ee

\noi and the solutions is

\be
F^{(\sigma )} =\tau ^{3/2} [c_1 H_\nu ^{(1)} ({|k_3 | b^2 (\tau )^{6}\over 3})+
c_2 H_\nu ^{(2)} ({|k_3 | b^2 (\tau )^{6}\over 3})],
\ee

\noi where

\be
\nu  ={ \sqrt{(3/2)^2 -(k_\perp /b)^2 }\over 3},
\ee

\noi  and $H_\nu ^{(i)} $ is a Hankel function of order
$\nu$.

\section{RESTRICTED SOLUTIONS}

In the previous section we considered those values of q for which it is
possible to solve
equation (10) for arbitrary values of $k_3$ and $k_\perp$, on the other hand
it is possible to solve the field equation for arbitrary values of q but the
particular case where $k_3$ or $k_\perp$ or both are zero.

\section {$k_3 =k_\perp= 0$ }
The field equation is in this case

\be
 {\ddot F}^{(\sigma)} +
\frac{1-3q}{\tau}\dot F^{(\sigma)}=0,
\ee
\noi The solutions are

\be
F^{\sigma}=\cases {c_1 +c_2 \tau ^{3q},& $q\ne 0$\cr
		    \cr
		    c_1 +c_2 \ln (\tau ),& $q=0$\cr}
\ee

\section{ $k_3 =0, k_\perp\ne 0$.  }

\noi The field equation and the solutions are

\be
 {\ddot F}^{(\sigma)} -2
{{\dot F}^{(\sigma)}\over \tau }+
[\frac{ k_\perp ^2}{ (b \tau )^2} ]    F^{(\sigma )}=0,
\ee

\be
F^{\sigma} =\cases{\tau ^{3q/2}[
c_1 Z_{\frac{q}{1-q}} (
\frac{2|k_\perp b^{\frac{1-3q}{2}}| \tau^{\frac{3(1-q)}{2}}}{{3(1-q)}}
		     )+c_2 Z_{\frac{-q}{1-q}} (
\frac{2|k_\perp b^{\frac{1-3q}{2}}| \tau^{\frac{3(1-q)}{2}}}{{3(1-q)}}
		     )];& $q\ne 1;$\cr
				   \cr
	      c_1\tau^\alpha + c_2\tau^\beta,& $q=1 ,   b^2\ne 4k_\perp^2/9$ \cr
	      \cr
	     \tau^\alpha (c_1+c_2 \log \tau ),&$q=1,   b^2= 4k_\perp^2/9$\cr}
\ee

where

\be
\alpha =\frac{3\pm  \sqrt{9 -4k_\perp^2/b^2}}{2} ,
\beta =\frac{3\mp  \sqrt{9 -4k_\perp^2/b^2}}{2}.
\ee

and $Z_\nu$ is a solution of Bessel equation of order $\nu$.

\section{ $k_3 \ne 0, k_\perp= 0$.  }

Now, Eq. (10) and its solutions are

\be
 {\ddot F}^{(\sigma)} +(1-3q)
{{\dot F}^{(\sigma)}\over \tau }+
[ k_3 ^2  (b \tau)^{6q-2} ]    F^{(\sigma )}=0,
\ee

\be
F^{\sigma} =\cases{\tau ^{3q/2}[
c_1 Z_{\frac{|q|}{2q}} (
\frac{|k_3 b^{{3q-1}}| \tau^{3q} }{{3q}}
		     )+c_2 Z_{\frac{|q|}{2q}} (
\frac{|k_3 b^{{3q-1}}| \tau^{3q} }{{3q}}
		     )]    ;& $q\ne 0;$\cr
				    \cr
	      c_1\tau^\alpha + c_2\tau^\beta,& $q=0 ,   b^2\ne 4k_\perp^2$ \cr
	      \cr
	     \tau^\alpha (c_1+c_2 \log \tau ),&$q=0,   b^2= 4k_\perp^2$\cr}
\ee

where

\be
\alpha =\frac{1\pm  \sqrt{1 -4k_\perp^2/b^2}}{2} ,
\beta =\frac{1\mp  \sqrt{1 -4k_\perp^2/b^2}}{2}.
\ee

and $Z_\nu$ is a solution of Bessel equation of order $\nu$.

In this paper we have found several exact solutions to the Maxwell equations
in
some anisotropic axisymmetric Bianchi type I cosmological models. The
possibility of having a self-consistent model for the
Einstein-Maxwell-Klein-Gordon equations, as well as the second
quantization and particle production is under consideration and will be
reported in a forthcoming paper.
\section {ACKNOWLEDGMENTS}

This work was supported by the CONACYT GRANT 1861-E9212.

\end{document}